\documentclass[twocolumn,prb,citeautoscript,superscriptaddress,8pt]{revtex4-2}

\usepackage{graphicx}
\usepackage{gensymb}
\usepackage{color}
\usepackage[dvipsnames]{xcolor}
\usepackage{float}
\usepackage{upgreek}
\usepackage{hyperref}
\hypersetup{urlcolor=blue, colorlinks=true, linkcolor=blue,citecolor=blue, linktocpage=true}
\usepackage{bm}
\usepackage{amsmath,amssymb}

\begin{document}

\title{Magnetic field imaging with an optical microscope using a quantum diamond sensor add-on}

\author{Alex Shaji} 
\affiliation{School of Science, RMIT University, Melbourne, VIC 3001, Australia}

\author{David A. Broadway} 
\affiliation{School of Science, RMIT University, Melbourne, VIC 3001, Australia}

\author{Philipp Reineck} 
\affiliation{School of Science, RMIT University, Melbourne, VIC 3001, Australia}

\author{Kevin J. Rietwyk} 
\email{kevin.rietwyk@rmit.edu.au}
\affiliation{School of Science, RMIT University, Melbourne, VIC 3001, Australia}

\author{Jean-Philippe Tetienne}
\affiliation{School of Science, RMIT University, Melbourne, VIC 3001, Australia}

\begin{abstract} 
Widefield magnetic imaging using ensembles of nitrogen-vacancy (NV) centres in diamond has emerged as a useful technique for studying the microscopic magnetic properties of materials.
Thus far, this technique has mainly been implemented on custom-made optical microscopes.
We have developed an add-on for a standard laboratory optical microscope that integrates the NV-diamond sensor and necessary light source, microwave antenna, and bias magnet, enabling NV-based magnetic imaging while retaining the typical optical measurements modes of the microscope. We demonstrate our retrofitted quantum diamond microscope by imaging a magnetic particle sample using brightfield, darkfield, and magnetic imaging modes. Furthermore, we employ an iso-magnetic field imaging technique to visualise the magnetic field of the sample within seconds, and finally demonstrate three-dimensional stray field imaging. Retrofitting existing microscopes exploits the stability and high quality of traditional optical microscope systems while reducing the cost and space requirements of establishing a standalone magnetic imaging system.
\end{abstract}

\maketitle

\section{Introduction}
The quantum diamond microscope (QDM) is a magnetic imaging technique which uses negatively charged nitrogen-vacancy (NV) centers in diamond as optically addressable quantum sensors to image stray magnetic fields emanating from samples. Due to the large field of view, high sensitivity and robustness of the NV center~\cite{Levine2019, Doherty2013}, the QDM has been used to investigate magnetic features in ancient magnetic fossils~\cite{Steele2023,Fu2020}, electronic devices~\cite{Turner2020,Kehayias2024,Garsi2024} and micromagnetic materials~\cite{Huang2023,Toraille2018,Kehayias2020,Mosavian2024,McCoey2020}. Quantum diamond microscopes are typically either large/bulky research laboratory setups for highly specialised measurements~\cite{Wang2024, Roncaioli2024, Xu2023} or bespoke table-top/portable systems ~\cite{Shaji2024, QDMIO2021}.
Either way, a limitation of these custom-made instruments is that they require their own laboratory space which may not be available to researchers interested in the capabilities of the QDM. 
However, as we have reported recently,~\cite{Abrahams2021,Shaji2024,Rietwyk2024} QDMs can be relatively straightforward to build since the technique itself only requires a few pieces of commercially available hardware and open source software for measurement and analysis~\cite{Volk2022_QDMlab, qudi-core, qdmpy, dukit}. The critical hardware components include a diamond with a thin layer of NVs (sensor chip), a light source to optically excite the NVs, an optical filter to remove the excitation light, a camera to image the photoluminescence (PL), and a tunable microwave source to drive the spin resonance of the NV.~\cite{Kim2019,Fu2024}. With this equipment it is straightforward to measure optically detected magnetic resonance (ODMR) spectra in order to perform magnetometry~\cite{Barry2020}. In fact, the optical components can be readily found in a typical fluorescence optical microscope while the remaining items can be added by mounting a diamond sensor chip on a PCB designed with integrated antenna driven by a microwave generator. Compared to a standalone QDM, adapting an existing optical microscope would improve the ease of use and access while enabling complementary measurements typical to optical microscopes and leveraging their inherent high mechanical stability and fine control. It would also drastically reduce the cost and remove the need to sacrifice precious laboratory space to establish QDM capabilities.

In this work, we present a 3D printed add-on for a commercial optical microscope that enables magnetic imaging. It slides around a single objective and integrates the diamond sensor chip, microwave source, magnet, and light source. The add-on allows access to the adjacent objectives in the turret for rapid switching between objectives and/or measurement modes. We demonstrate the capabilities of the augmented optical microscope by measuring magnetic microparticles using brightfield, darkfield, and magnetic imaging modes. We present two different magnetic imaging approaches, full ODMR and iso-B for fully quantitative and real-time imaging, respectively. We further demonstrate three-dimensional stray field imaging by tuning the distance between the diamond and sample using the existing high-stability mechanical stage of the optical microscope. Our work demonstrates that retrofitting existing optical microscopes to enable QDM magnetic imaging is possible, which may facilitate a broader adoption of QDM techniques.

\section{Results and Discussion}
The add-on assembly for retrofitting an optical microscope we develop in this work is shown in Fig.~\ref{fig1}(a) and a CAD file is provided in the Supplementary Information. It is comprised of three main parts: (1) a baseplate, (2) a collar and (3) a PCB holder -- each will be described in detail below. 
(1) The baseplate has two arms and 25\,mm clear aperture lens tube ($\approx 10$\,mm long) with external threading (2\,mm pitch) in which the objective sits -- the aperture at the microscope end is narrower than the objective such that screwing the objective into the microscope clamps the base plate. One of the base plate arms consists of a hinge and a Thorlabs 16\,mm cage plate to allow mounting of a fibre adapter and collimating lens via $1/2\,''$ lens tube. 
A multi-mode fibre delivers light from a 532\,nm continuous wave (CW) laser and the output is collimated/focused at the diamond with an achromatic lens (30\,mm focal length). 
The other arm consists of a curved rail centred around the focal point of the objective, which is used for mounting and tuning the angle of a permanent magnet.
(2) A collar with a freely spinning nut is used to adjust the vertical height of the diamond sensor and bring the sensor into focus while allowing the collar and diamond to rotate in order to align the diamond NV axes to the applied magnetic field. 
(3) A custom PCB consisting of an 8\,mm microwave (MW) loop antenna is glued to the legs of the PCB holder and connected to the collar via dovetails for easy mounting and some lateral adjustment. The diamond sensor (with dimensions 4\,mm x 4\,mm x 400\,$\mu$m) was glued to a piece of glass slide and attached to the PCB. 
Photographs of the realised add-on are shown in Fig.~\ref{fig1}(b), which is mounted to a standard commercial optical microscope (ProSciTech OXJS304). Here the add-on is attached to the Plan Achromatic 4x, numerical aperture NA = 0.10 objective supplied with the microscope. Although the add-on is designed for an objective lens with a 24\,mm diameter and parfocal length of 45\,mm it can easily be adjusted to accommodate other objectives. The PL from the diamond is collected using the objective and imaged onto the camera using the optics built in the microscope. Here we attach our own camera (Basler acA2040-90um USB3 Mono) but in principle any pre-installed camera can be used. An emission filter (650\,nm longpass) must be inserted before the camera to block the laser light; in future iterations, the filter could be integrated into the add-on, for instance between the diamond and objective. 
The diamond used in this study is a high-pressure high-temperature (HPHT) grown single crystal substrate with a 500\,nm NV layer \cite{Abrahams2021}. In addition to the add-on, there is a control box described elsewhere \cite{Shaji2024} which comprises a microwave (MW) signal generator (Windfreak SynthNV Pro), an ampliﬁer (Mini-Circuits ZHL-42), and the 532\,nm laser (Laser Quantum Opus 2W). A custom LabVIEW program is used for recording optical images and performing QDM measurements, while a Python script based on QDMpy.~\cite{qdmpy} is used for further analysis and post-processing. Detailed descriptions of the PCB design, microwave control box and the LabVIEW and Python codes are provided in Ref. ~\cite{Shaji2024}.

\begin{figure}[t!]
    \centering
    \includegraphics[width=8.5cm]{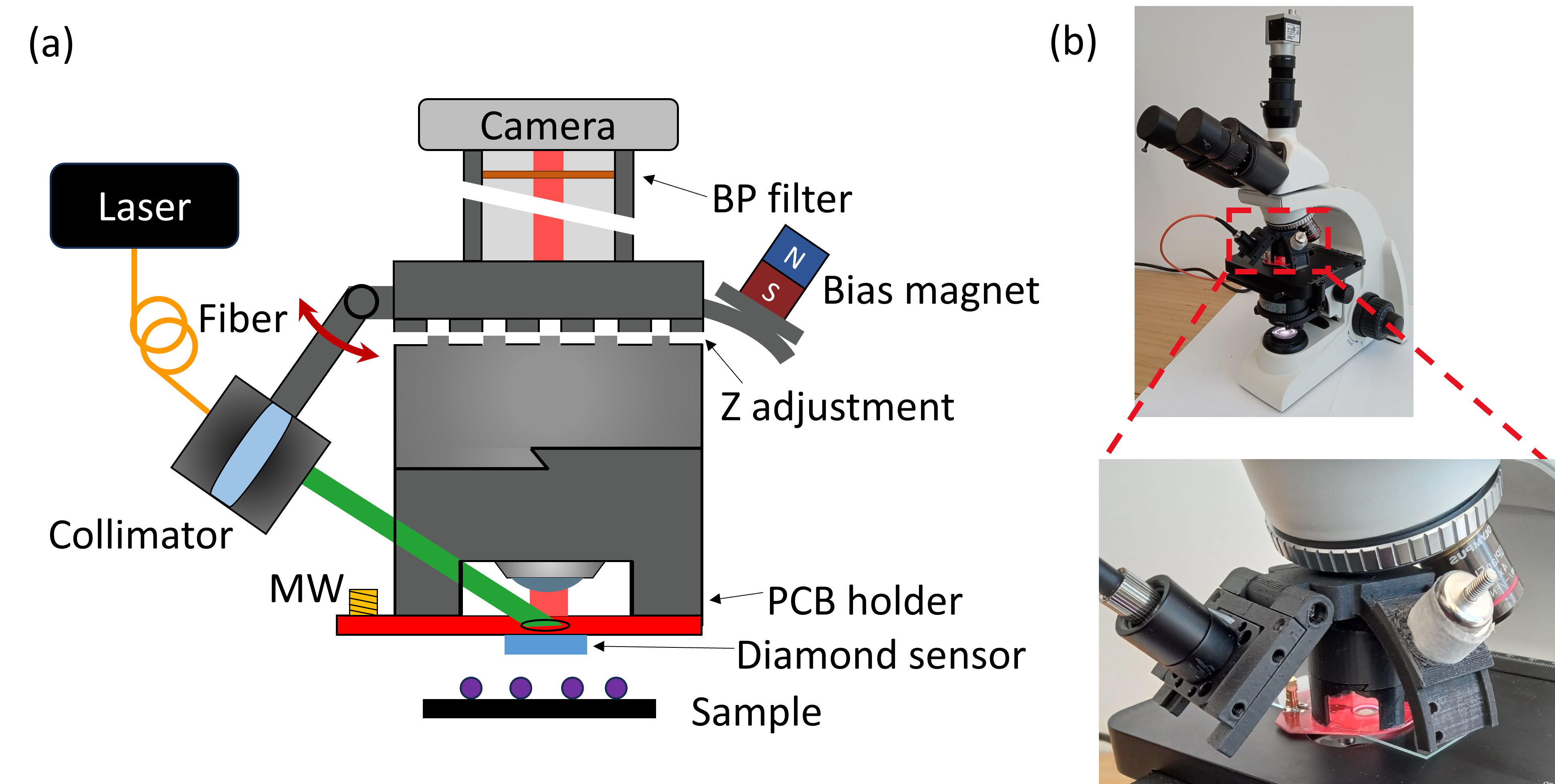}
    \caption{(a) Schematic of a 3D printed add-on consisting of a PCB-diamond assembly, MW antenna, light source collimation, and bias magnet. The add-on attaches to the turret of a commercial optical microscope to enable QDM imaging. (b) Photographs of the add-on attached to a 4x objective on a ProSciTech OXJS304 trinocular microscope. The other objectives on the turret remain available for standard optical imaging.}
    \label{fig1}
\end{figure}
\begin{figure}[t!]
    \centering
    \includegraphics[width=8.5cm]{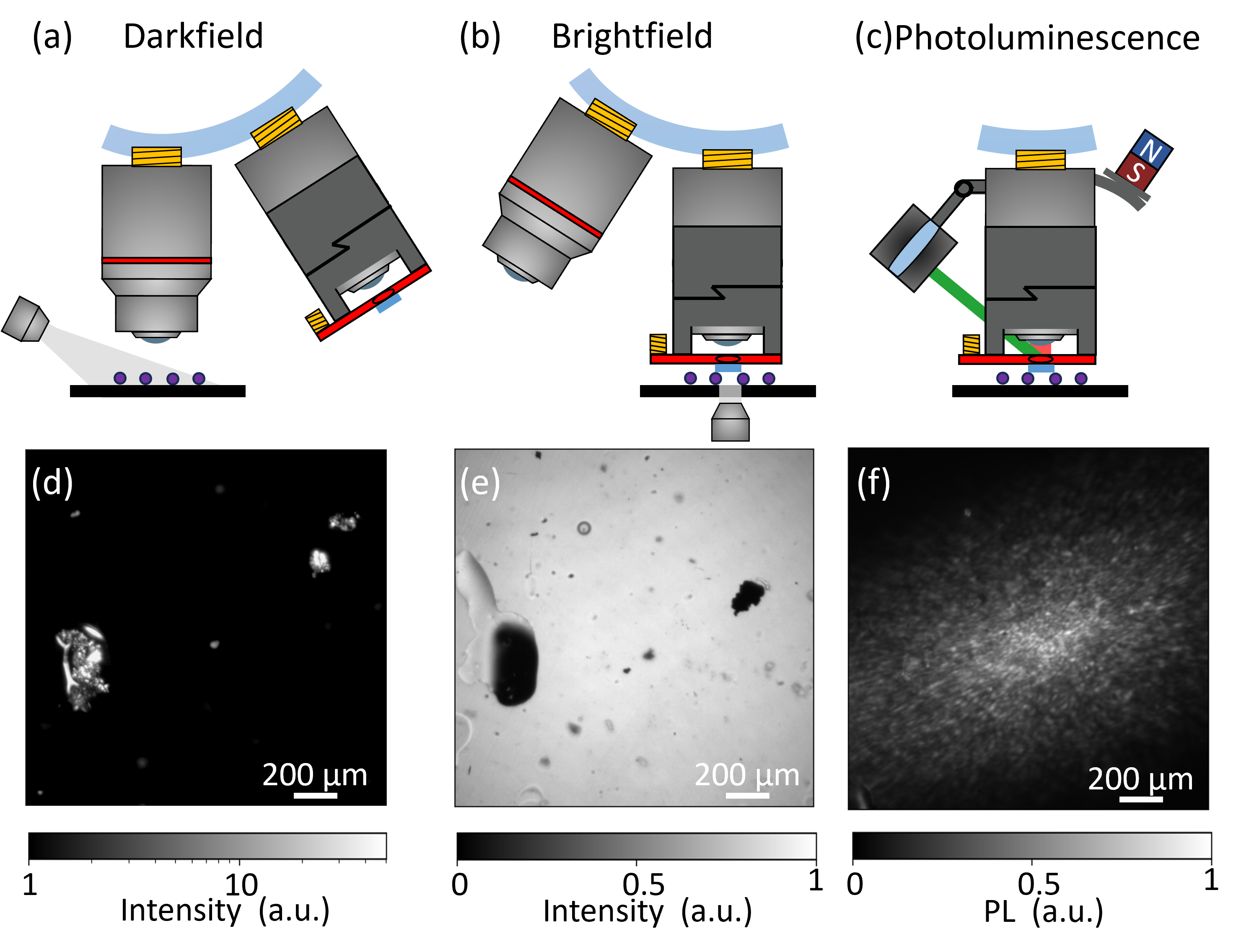}
    \caption{
    (a-b) Schematics showing two 4x objectives on a microscope turret, one with our add-on, and a sample (two main fragments from a Neodymium magnet glued to a glass coverslip). In (a), darkfield imaging is achieved using side illumination and a bare objective. 
    In (b),  brightfield transmission through the bulk diamond substrate is achieved with white light from below. 
    (c) For PL imaging, the fibre-coupled laser delivered by the add-on is turned on.
    (d,e) Widefield images obtained using a 4x objective, (d) Darkfield image without the add-on and (e) brightfield transmission image with the add-on assembly. 
    (f) PL image of the NV layer across the same region of the sample.}
    \label{fig2}
\end{figure}

To demonstrate the effectiveness of our QDM add-on, we measure some magnetic particles using traditional measurements modes of the optical microscope including darkfield, brightfield transmission, and photoluminescence, before performing magnetic imaging later in this work. We use microparticle fragments from a Neodymium (Nd) magnet glued onto a thin glass coverslip as a test sample. For darkfield measurements we use a fibre illuminator to shine white light on the sample from the side, as illustrated in Fig. \ref{fig2}(a). Such images are useful for locating microparticles and choosing a region of interest (ROI), Fig. \ref{fig2}(d). 
Next, we turn the turret onto a 4x objective with the attached add-on and perform transmission imaging of the sample using the built-in white light source as shown in Fig. \ref{fig2}(b,e). These brightfield transmission measurements show all the features present in the sample and on the diamond surfaces and are useful for tuning the optical focus of the diamond sensor-sample interface. Fortunately, most of these are non-magnetic and will not appear in magnetic images. To generate magnetic images, the laser (210\,mW at the sample) is turned on to excite the PL from the NV layer in the diamond (Fig. \ref{fig2}(c)) -- a typical PL image used to form the magnetic field image is shown in Fig. \ref{fig2}(f). The elliptical shaping is spreading of the laser spot caused by the grazing angle illumination. Centering of the laser spot is achieved by simple adjustment of the arm holding the collimating lens. As seen in Fig. \ref{fig2}(f), the speckling of the laser prevents easily observing features in the sample and highlights the need for other measurement modes and light sources. 

\begin{figure}[t!]
    \centering
    \includegraphics[width=8.5cm]{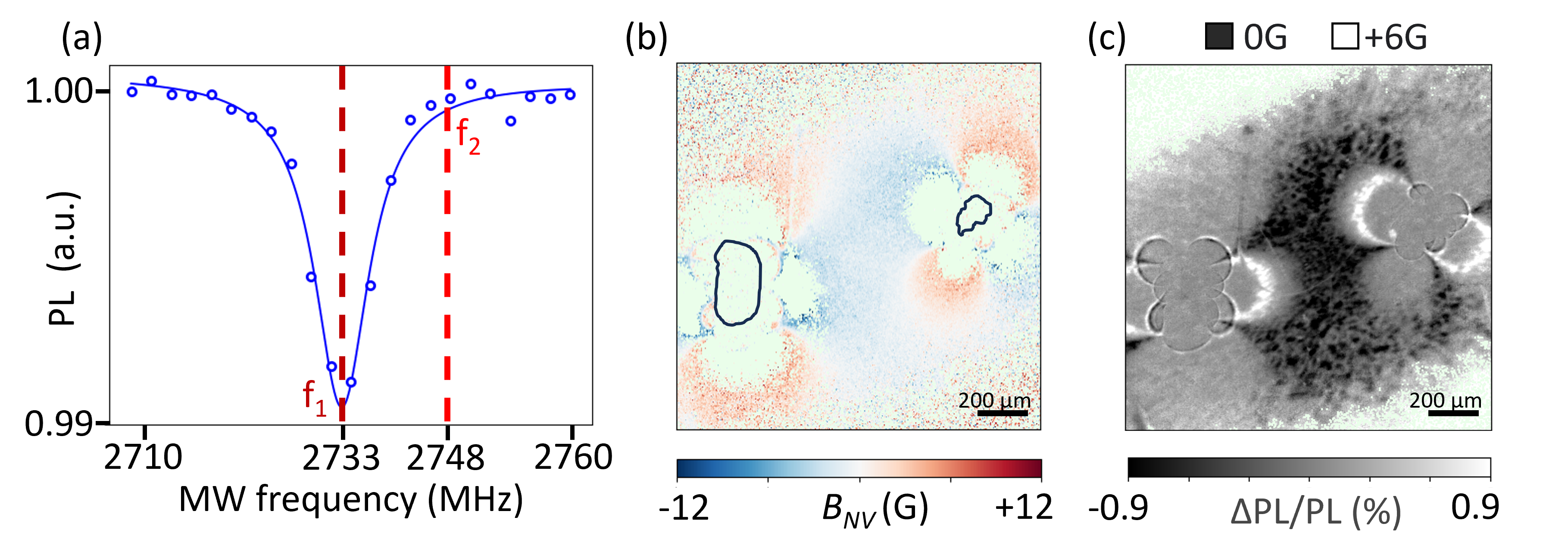}
    \caption{(a) Single peak ODMR spectrum integrated over a small region in (b). This spectrum data is fitted with a single Lorentzian peak. (b) $B_{\rm NV}$ field map of magnetic microparticles over the region corresponding to the PL image in Fig. \ref{fig2}(f). (c) Asymmetric iso-B field image (integration time of 216\,s) showing the dark and bright contours of the stray field $B_{\rm NV}$ = 0 and 6 G respectively.} 
    \label{fig3}
\end{figure}

Now we consider magnetic imaging of the microparticles using our add-on. To begin, the diamond surface is brought into focus by rotating the nut on the collar and the PCB holder is rotated to ensure that one of NV axes of the diamond is aligned with the applied magnetic field. The sample is brought into contact with the sensor using the microscope stage. A CW ODMR spectrum is obtained by recording PL images as the MW frequency is swept and this sweep is repeated and the images integrated to improve the signal-to-noise. In order to minimise the integration time, we measured only one of the two peaks corresponding to the NVs aligned to the magnetic field. An example spectrum taken over a small region of interest devoid of magnetic features is shown in Fig. \ref{fig3}(a). To acquire a magnetic field map, the ODMR spectrum at each pixel is fit with a Lorentzian distribution to obtain the peak position $f$ and the corresponding magnetic field is calculated using $B({\rm G})=\frac{f({\rm MHz})-2870}{2\gamma_e}$ where $\gamma_e=2.80$\,MHz/G is the electron gyromagnetic ratio. The applied field (48.9\,G) is subtracted from this map to obtain the stray magnetic field projected along the NV axis ($B_{\rm NV}$) -- an example of a stray field map of the microparticles is shown in Fig. \ref{fig3}(b) with an outline of each microparticle determined from the brightfield image in Fig. \ref{fig2}(e). Where the magnetic field strength (close to the microparticles) exceeds $\pm$ 12\,G, the corresponding frequency shifts in the ODMR spectrum exceed the measured range and the fit $B_{\rm NV}$ is unreliable. Such regions are shaded in pale green for clarity and limit the spatial information that can be obtained from the map. The integration time for the magnetic field measurements, to obtain the desired signal-to-noise ratio (SNR), was 45\,minutes. To better complement typical widefield optical microscopy techniques which require only seconds, a faster analysis protocol would be beneficial.

In order to address the limited spatial information from the magnetic maps and to optimise measurement time, we demonstrate iso-magnetic (iso-B) imaging mode, a technique that has been used to image various magnetic materials~\cite{Rondin2013,Rondin2012,Gross2017,Maletinsky2012}. 
The difference in PL ($\Delta PL$) measured at two MW frequencies $f_{1}$ and $f_{2}$ (marked in Fig. \ref{fig3}(a)) normalised to PL($f_{1}$) is mapped across every pixel of the image as shown in Fig. \ref{fig3}(c). 
The MW frequencies of $f_{1}$ = 2733 and $f_{2}$ = 2748 MHz were chosen, corresponding to fields of $B_{\rm NV}$ = 0 and 6\,G, shown as dark and bright contours respectively. The bright contours reveal the direction of the fields from the microparticles while dark contour show that between the two particles the fields are repelling and there is a region with near to no field. Regions where the PL intensity is low due to limited excitation intensity, notably, the top-left and bottom-right corners (see Fig. \ref{fig2}(f)), have been masked and shaded pale green.

\begin{figure}[t!]
    \centering
    \includegraphics[width=8.5cm]{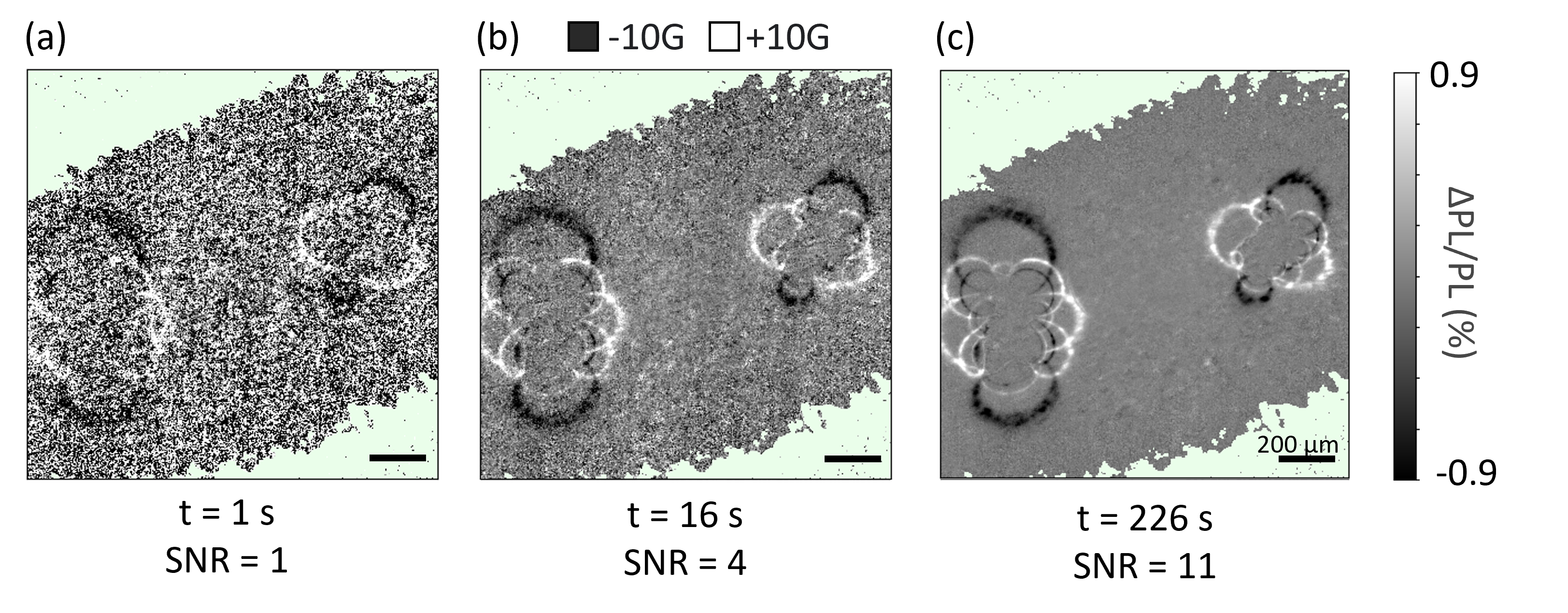}
    \caption{(a-c) Dual iso-magnetic field images where dark and bright contours represent where the stray field projected along the NV axis $B_{\rm NV}$ = -10 and 10\,G respectively with integration times (a) 1, (b) 16 and (c) 226\,s.}
    \label{fig4}
\end{figure}

The frequencies chosen for iso-B imaging can be varied to produce different contours. For symmetric dual iso-B imaging, we tune $f_{1}$ and $f_{2}$ to $\pm28$\,MHz frequency shifts from the resonant frequency (2705 and 2761\,MHz) corresponding $B_{\rm NV}$ = -10 and +10\,G and in Fig. \ref{fig4}(a-c) we show the evolution in the magnetic images as a function of time. The critical change is an improvement in SNR with time. After 1\,s integration, there are clear bright and dark contour lines indicating that this analysis can be used for real-time, rapid screening and/or visualisation of magnetic features. With further integration the resolved contour lines become finer and more pronounced due to reduction in the noise. The total integration time of 226\,s is a fraction of the time required for a full magnetic field image (tens of minutes to hours) since the PL is considered at only two frequencies. SNR values shown for images in Fig. \ref{fig4}(a-c) are calculated by dividing mean signal by standard deviation (noise) in those images across a small region.

\begin{figure}[t!]
    \centering
    \includegraphics[width=8.5cm]{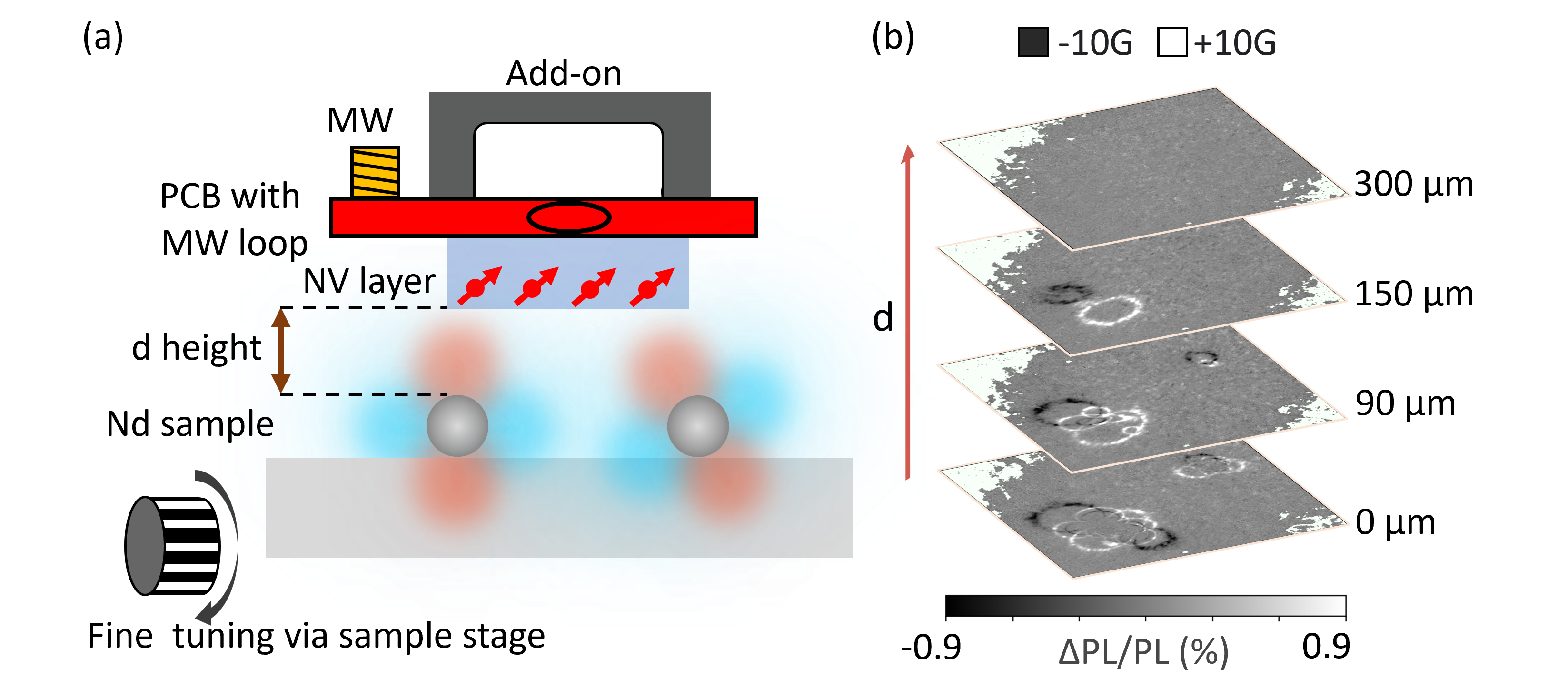}
    \caption{(a) Schematic for performing stray magnetic field tomography by fine tuning sensor distance. (b) Dual iso-magnetic field images for sample-sensor distances shown as 300, 150, 90 and 0\,$\mu$m (top to bottom).}
    \label{fig5}
\end{figure}

By combining the rapid analysis of iso-B imaging with the built-in fine sample stage control in the z-direction of the microscope, we can perform a simple tomography analysis. Starting with the diamond in contact with the sample, the sample-diamond sensor separations $d$ can be varied with regular iso-B measurements, as illustrated in Fig. \ref{fig5}(a). The field map at $d = 0\,\mu$m reveals sharp features (Fig. \ref{fig5}(b). With increasing $d$, the magnetic field lobes gradually shrink until no features are visible at $d$ = 300\,$\mu$m. Magnetic field maps at different vertical displacements can be used to build up a 3D perspective of the intrinsic magnetic profiles of the sample under test and may be used to construct a tomography image of dual contours. The ability to rapidly and easily tune $d$ is useful for studying the magnetic moment of paleomagnetic samples with varying degrees of dipolarity at at the optimal height – $d$ is increased to ensure sufficient integration of the magnetic features of a grain at the compromise of the moment sensitivity~\cite{Fu2020}.

Having demonstrated a number of measurements modes and features offered by our add-on, it is worthwhile to consider possible improvements. 
The highest possible spatial resolution in magnetic imaging is limited by the separation of the magnetic features and the NV layer due to spreading out of the magnetic field with distance which for parallel surfaces can be restricted by the tilt between the surfaces. Although not an issue here because of the size of the particles, for other samples the incorporation of two perpendicular goniometers to finely tune the diamond surface tilt relative to the sample \cite{Rietwyk2024} may prove to be effective in achieving high resolution imaging with the add-on. 
Additionally, cylindrical lenses could be incorporated into the optics assembly for tuning the laser spot size and improving the uniformity of the spot, depending on user requirements.  
The current implementation of iso-B measures two magnetic contours but in principle more frequencies could be measured in order to produce a more detailed contour map with a compromise between speed and precision.

\section{Conclusion}
To summarise, we presented a 3D printed retrofit accessory for traditional optical microscopes that integrates the required components to enable magnetic imaging. Using a test sample of Nd magnetic microparticles we demonstrate standard widefield imaging modes with one objective, unobstructed by the diamond chip, before switching to the objective with our add-on attached and performing magnetic imaging. Two different magnetic mapping methods were used, ODMR spectroscopy for high accuracy and iso-B for real-time analysis that is complementary to the other widefield techniques. Due to the ubiquity of optical microscopes in research laboratories, the add-on, control box and laptop constitute an accessible implementation of the QDM which requires minimal additional laboratory space. 


\begin{acknowledgments}
This work was supported by the Australian Research Council (ARC) through grants FT200100073, DP220100178 and DE230100192. 
P.R. acknowledges support through an RMIT University Vice-Chancellor’s Research Fellowship. 
\end{acknowledgments}

\section*{Data Availability Statement}

The data that supports the findings of this study are available within the article and its supporting information.

\section*{Author Declarations}

The authors have no conflicts to disclose.

\bibliography{refs}

\end{document}